\documentclass[aps,prl,groupedaddress,showpacs,tighten]{revtex4}
\usepackage{graphicx}
\usepackage{times}
\def\beq{\begin{equation}}
\def\eeq{\end{equation}}
\def\bey{\begin{eqnarray}}
\def\eey{\end{eqnarray}}

\def\lsim{\mathrel{\raise.3ex\hbox{$<$\kern-.75em\lower1ex\hbox{$\sim$}}}}
\def\gsim{\mathrel{\raise.3ex\hbox{$>$\kern-.75em\lower1ex\hbox{$\sim$}}}}

\begin{document}

\title{Challenges in Detecting Gamma-Rays From Dark Matter Annihilations in the Galactic Center}  
\author{Gabrijela Zaharijas and Dan Hooper}
\address{Fermi National Accelerator Laboratory, Particle Astrophysics Center, Batavia, IL  60510}

\date{\today}

\begin{abstract}

Atmospheric Cerenkov Telescopes, including HESS and MAGIC, have detected a spectrum of gamma-rays from the galactic center region which extends from $\sim$200 GeV or lower, to at least $\sim$10 TeV. Although the source of this radiation is not yet known, the spectrum appears to behave as a simple power-law, which is not the expectation for gamma-rays generated through the annihilation of dark matter particles. If instead we conclude that the source of these gamma-rays is astrophysical in origin, this spectrum will constitute a background for future dark matter searches using gamma-rays from this region. In this paper we study how this background will affect the prospects for experiments such as GLAST to detect dark matter in the galactic center. We find that only a narrow range of dark matter annihilation rates are potentially observable by GLAST given this newly discovered background and considering current constraints from EGRET and HESS. We also find that a detection of line emission, while not completely ruled out, is only possible for a very narrow range of dark matter models and halo profiles. 
\end{abstract}
\pacs{95.35.+d;95.30.Cq,95.55.Ka; FERMILAB-PUB-06-048-A}
\maketitle

\section{Introduction}

Many methods of detecting weakly interacting, massive dark matter particles have been explored~\cite{review}. In addition to collider searches, direct and indirect astrophysical techniques are being actively pursued. Direct detection experiments attempt to observe the elastic scattering of dark matter particles in a detector, whereas indirect detection consists of observing the annihilation products of dark matter particles, such as neutrinos, positrons, anti-protons, anti-deuterons, or gamma-rays.

Indirect dark matter searches using gamma-ray telescopes, unlike anti-matter experiments, have the advantage of being able to search for point sources of annihilation radiation. Nearby regions with very high densities of dark matter make for the brightest of such point sources. The center of our galaxy has long been considered to hold the greatest prospects for observing gamma-rays from dark matter annihilations, particularly if the halo profile of the Milky Way is cusped in its inner volume \cite{gchist,dingus,buckley}.

Recently, observations by four Atmospheric Cerenkov Telescopes (ACTs), HESS \cite{hess}, WHIPPLE \cite{whipple}, MAGIC \cite{magic} and CANGAROO-II \cite{cangaroo}, have detected a very bright gamma-ray source in the direction of the galactic center. The spectrum of this source has been measured in increasing detail, in particular by HESS \cite{hess}. Although the first HESS data from this source appeared to be marginally consistent with a spectrum from annihilating dark matter \cite{hessdark,kkbergstrom,actdark,heavyneutralino}, as more data has been accumulated, it is now difficult to reconcile the HESS data with such a spectrum. The $E^{-2.2}$ power-law spectrum measured by HESS, which extends over nearly two decades of energy, appears more likely to have been generated by an astrophysical accelerator. Additionally, to produce gamma-rays in the energy range observed by HESS, a dark matter candidate would have to be very heavy ($\sim$10-50 TeV), well beyond the natural range for neutralinos and other well motivated dark matter candidates (for an exception, see Ref.~\cite{hessdark}).

The presence of a bright astrophysical source in the galactic center region would pose a serious challenge to future dark matter searches. In particular, the prospects of GLAST being able to detect the presence of dark matter in the galactic center will be considerably diminished as a result of the presence of the HESS source.  In this article, we study in detail the impact of the HESS source on future dark matter searches.

\section{The TeV Gamma-Ray Source at the Galactic Center}

The center of our galaxy is a rich and complex region. It includes a $2.6\times 10^6\, M_{\odot}$ black hole, coincident with the radio source Sgr A$^*$, which has also been observed to produce variable emission at infra-red and X-ray wavelengths. Only a few parsecs from Sgr A$^*$ is the supernova remnant Sgr A East. In addition, this region contains a wide variety of notable astrophysical objects, including massive O and B type stars and massive compact star clusters (Arches and Quintuplet). 

The gamma-ray spectrum from the galactic center (coincident with Sgr A$^*$), as reported by the HESS collaboration, has been measured at energies between $\sim 200$ GeV and $\sim 10$ TeV. This lower value is a result of HESS's energy threshold, while the upper value is only a matter of limited data due to a steadily falling spectrum. This data is fit quite well by a simple power-law spectrum, $dN_{\gamma}/dE_{\gamma} \propto E^{-2.2}_{\gamma}$ \cite{hess}. Although the source of this radiation has not been identified, it is not difficult to imagine such a spectrum being generated by an astrophysical accelerator. The spectrum resulting from dark matter annihilations, on the other hand, fits the observed spectrum considerably less well. The spectrum of gamma-rays produced in the cascade decays of heavy quarks, leptons or gauge bosons generated in dark matter annihilations, in contrast, does not follow a simple power-law, but instead features a continuously changing slope. The constant slope power-law spectrum measured by HESS does not appear to be consistent with annihilating dark matter.

Astrophysical origins for the TeV source at the galactic center have also been proposed. In particular, it has been suggested that this emission may be generated through acceleration processes associated with our galaxy's central supermassive black hole \cite{ahro,dermer}. Unlike other known galactic and extragalactic accelerators associated with black holes, however, Sgr A$^{\star}$ has a very low bolometric luminosity, which allows for very high energy gamma-rays to escape the source, whereas they would likely be absorbed in the dense radiation fields surrounding a more luminous accelerator.

\begin{figure}

\resizebox{10cm}{!}{\includegraphics{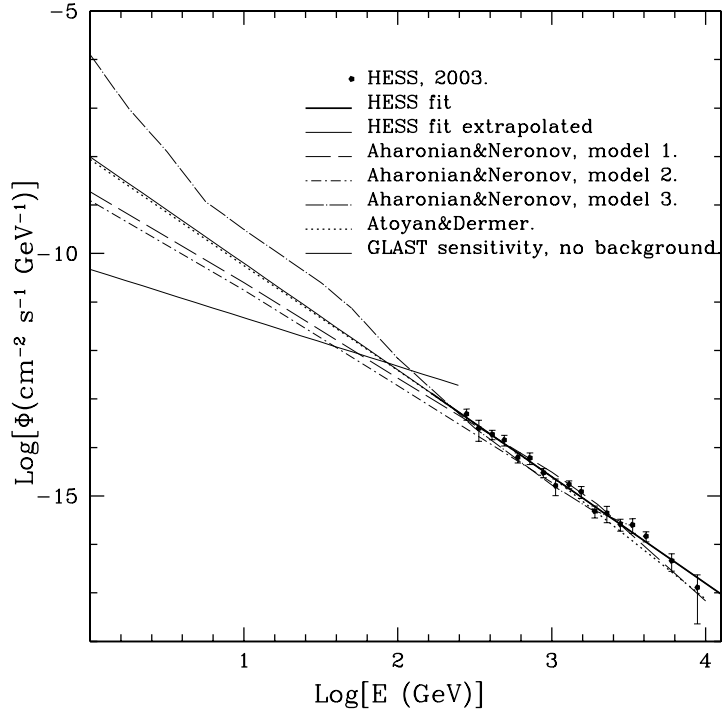}}
\caption{The gamma-ray spectrum observed by HESS (points and error bars) and the spectrum predicted by several astrophysical models \cite{ahro,dermer}. Also shown is the spectrum as extrapolated from the HESS data, as well as the projected sensitivity of GLAST when the background is neglected (95\% CL, following Ref.~\cite{stat}). See text for more details.}
\label{astro}
\end{figure}

In Fig.~\ref{astro}, we show the HESS data along with the spectra of several astrophysical models which match the HESS data reasonably well in the measured energy range. The model shown involving photo-meson interactions of ultra-high energy protons with ambient photons and magnetic fields (Aharonian and Neronov, model 1~\cite{ahro}), as well as the model of inelastic proton-proton collisions of multi-TeV protons in the accretion disk (Aharonian and Neronov, model 2~\cite{ahro}), and the model of Atoyan and Dermer \cite{dermer} each predict a spectrum which behaves largely as a continuous power-law over the energy range shown. This is not the case, however, for the spectrum resulting from curvature and inverse Compton radiation (Aharonian and Neronov, model 3~\cite{ahro}), which deviates considerably from a simple power-law below $\sim$100 GeV.

\section{Limits on the Dark Matter Annihilation Rate From HESS}

\begin{figure}

\resizebox{8cm}{!}{\includegraphics{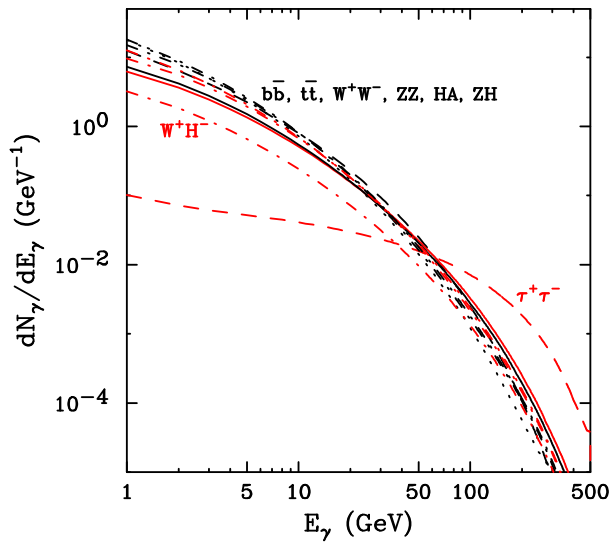}}
\caption{The gamma-ray spectrum (per annihilation) generated through dark matter annihilations for a variety of channels. A 500 GeV dark matter mass has been used in this example.}
\label{dmspec}
\end{figure}

To begin, we will determine the maximum annihilation rate of dark matter in the galactic center which is consistent with the observations of HESS. The spectrum of gamma-rays generated in dark matter annihilations is given by:
\begin{equation}
\Phi_{\gamma}(E_{\gamma},\psi) = <\sigma v> \frac{dN_{\gamma}}{dE_{\gamma}} \frac{1}{4\pi m^2_{\rm{dm}}} \int_{\rm{los}} \rho^2(r) dl(\psi) d\psi,
\label{flux1}
\end{equation}
where $<\sigma v>$ is the dark matter self-annihilation cross section (multiplied by velocity), $\psi$ is the angle away from the direction of the galactic center that is observed, $\rho(r)$ describes the dark matter density profile of the inner galaxy, and the integral is performed over the line-of-sight. $dN_{\gamma}/dE_{\gamma}$ is the gamma-ray spectrum generated per annihilation. This can vary depending on the annihilation mode considered. In Fig.~\ref{dmspec}, we show $dN_{\gamma}/dE_{\gamma}$ for various annihilation channels. In most cases ($b \bar{b}$, $t\bar{t}$, $W^+W^-$, etc.), the spectrum produced does not vary much. The exception to this is the somewhat harder spectrum generated through annihilations to $\tau^+ \tau^-$. Particle physics models very rarely predict a dark matter candidate that annihilates mostly to $\tau^+ \tau^-$, however. Neutralinos, for example, generally annihilate through this channel only a few percent of the time or less, although carefully selected model parameters can allow for this to be exceeded. We will assume throughout our study that the continuum gamma-ray emission from dark matter annihilation resembles the spectrum shown in Fig.~\ref{dmspec} for heavy quarks or gauge bosons, etc., rather than $\tau^+ \tau^-$.

The HESS source is consistent with a point source, with any extension of more than 3 arcminutes (0.05$^{\circ}$) conflicting with the data~\cite{hess}. Such point-like behavior is consistent with either an astrophysical source or dark matter annihilation in the case of a strongly cusped or spiked halo profile. For a standard NFW profile \cite{nfw}, for example, with $\rho(r) \propto r^{-1}$ in the inner halo, extension near the level of the HESS limit would be expected in the angular distribution of dark matter annihilation radiation \cite{actdark}. Less extension would be found for a Moore {\it et al.} profile ($\rho(r) \propto r^{-1.5}$) \cite{moore}, or for a density spike \cite{spike}, for example.

If the dark matter halo profile in the inner galaxy is sufficiently steep (roughly $\rho \propto r^{-1}$ or steeper), it will appear as a point source within the current level of angular resolution and we can rewrite Eq.~\ref{flux1} as a quantity averaged over a solid angle, $\Delta \Omega$:
\begin{equation}
\Phi_{\gamma}(E_{\gamma}) \approx 2.8 \times 10^{-12} \, \rm{cm}^{-2} \, \rm{s}^{-1} \, \frac{dN_{\gamma}}{dE_{\gamma}} \bigg(\frac{<\sigma v>}{3 \times 10^{-26} \,\rm{cm}^3/\rm{s}}\bigg)  \bigg(\frac{1 \, \rm{TeV}}{m_{\rm{dm}}}\bigg)^2 J(\Delta \Omega) \Delta \Omega,
\label{flux2}
\end{equation}
where the quantity $J(\Delta \Omega)$ depends only on the dark matter distribution, and is the average over the solid angle of the quantity:
\begin{equation}
J(\psi) = \frac{1}{8.5 \, \rm{kpc}} \bigg(\frac{1}{0.3 \, \rm{GeV}/\rm{cm}^3}\bigg)^2 \, \int_{\rm{los}} \rho^2(r) dl(\psi) d\psi.
\end{equation}

\begin{figure}

\resizebox{8cm}{!}{\includegraphics{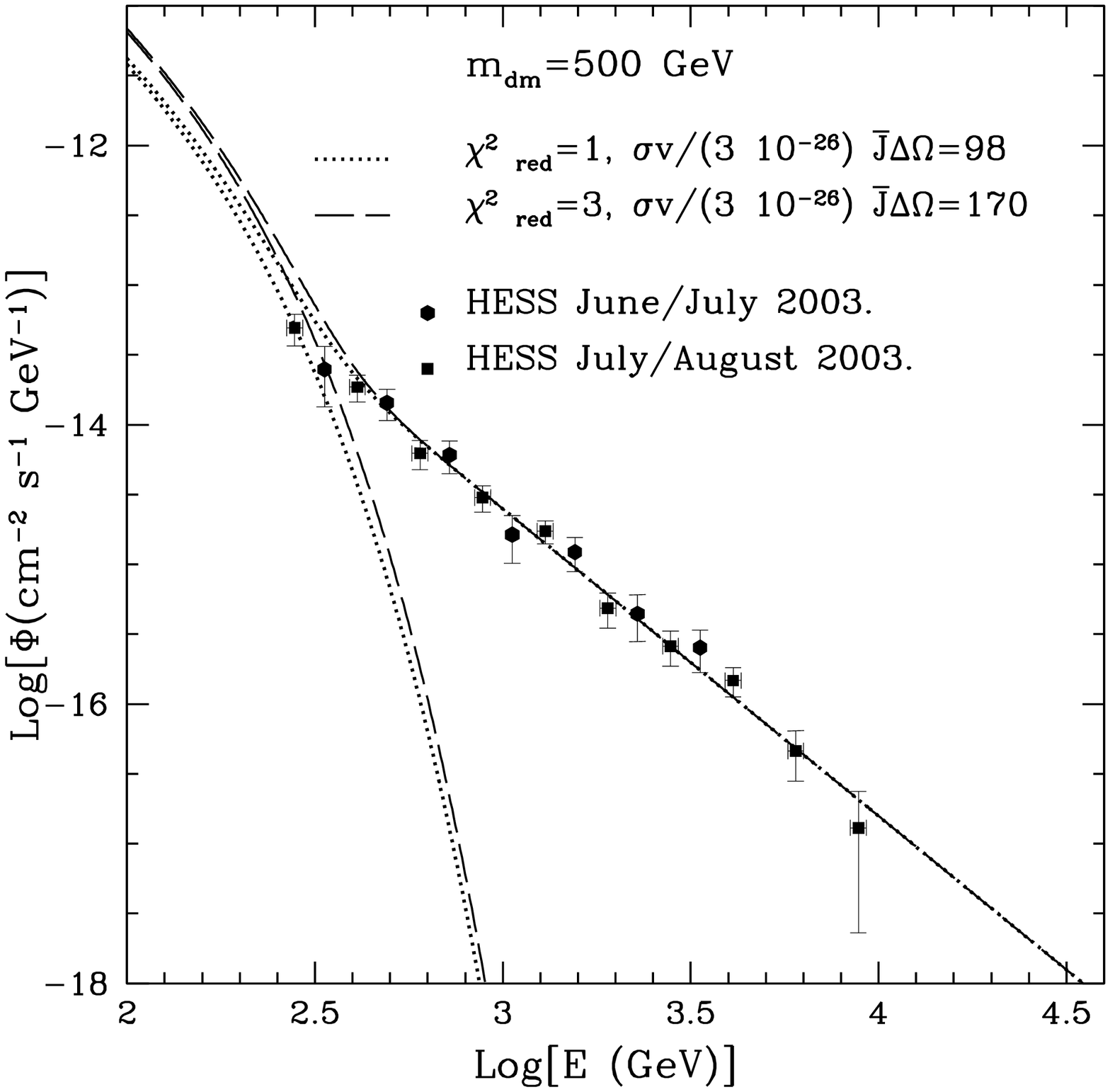}}
\resizebox{8cm}{!}{\includegraphics{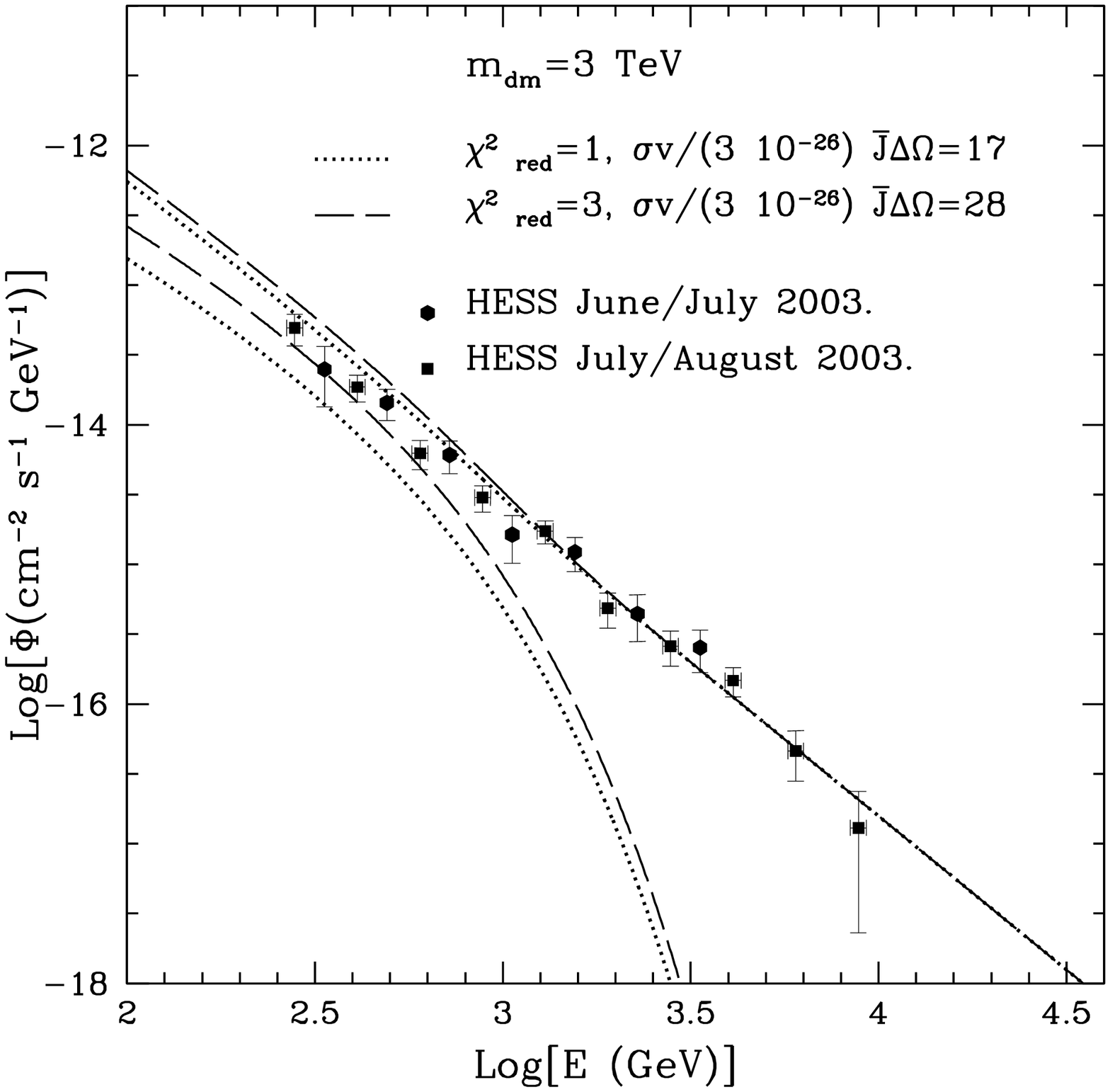}}
\caption{The maximum flux of gamma-rays from dark matter in the galactic center consistent with the HESS data. The two frames correspond to two dark matter masses: 500 GeV (left) and 3 TeV (right). For a 500 GeV WIMP, the quantity $(<\sigma v>/10^{-26}$cm$^3$/s)$ \times J(\Delta \Omega) \Delta \Omega$ can be as large as $\sim$100 without exceeding the HESS data. For a 3 TeV WIMP, this quantity can be as large as $\sim$20.}
\label{hess}
\end{figure}

In Fig.~\ref{hess}, we show the maximum gamma-ray flux which can be produced through dark matter annihilations in the galactic center without conflicting with the observations of HESS. Results are shown for two WIMP masses (500 GeV and 3 TeV). Adding a gamma-ray spectrum from dark matter annihilations to an astrophysical power-law background, we have increased the dark matter annihilation rate until the fit to the data becomes poor. In particular, we determine that the quality of the fit to the HESS data becomes poor ({\it ie.} the $\chi^2$ per degree of freedom exceeds 3) for values of $(<\sigma v>/10^{-26}$cm$^3$/s) $\times J(\Delta \Omega) \Delta \Omega$ larger than approximately 100 and 20, for a 500 GeV and 3 TeV WIMP, respectively. Lighter WIMPs are far less constrained by the HESS data.

The values of $<\sigma v>$ and $J(\Delta \Omega) \Delta \Omega$ can vary substantially from model to model. A WIMP which freezes out thermally in the early universe to generate the observed dark matter abundance can have a low velocity annihilation cross section as large as $\approx 3 \times 10^{-26}$ cm$^{-3}$/s, although smaller values are possible if terms proportional to $v^2$ dominated during freeze-out. A non-thermal dark matter relic could have a larger cross section as well. N-body simulations suggest that the inner regions of dark matter halos will be cusped, with a distribution which behaves as $\rho(r) \propto r^{-\gamma}$, where $\gamma$ generally lies in the range of 1 to 1.5. The NFW \cite{nfw} and Moore {\it et al}. \cite{moore} profiles are two of the most well known cusped profiles based on N-body results. Over solid angles of $\sim 10^{-4}$ sr, such profiles lead to  $J(\Delta \Omega) \sim 10^4$ to $10^6$.

There are good reasons to doubt the validity of N-body simulations in this application, however. Firstly, the resolution of such simulations is limited to scales on the order of a kiloparsec, whereas we are most interested in the dark matter distribution in the innermost few parsecs of our galaxy. It is not clear that extrapolating over this range will prove reliable. Secondly, the gravitational potential in the inner region of the Milky Way is dominated not by dark matter, but by baryons, which are not included in N-body simulations. The effect of baryons on the dark matter distribution is not well understood, however, and could plausibly result in either an enhancement or a reduction in the dark matter annihilation rate \cite{ac}. Furthermore, the adiabatic accretion of dark matter onto the central supermassive black hole may lead to the formation of a density spike in the dark matter distribution. The existence of such a spike would result in a very high dark matter annihilation rate \cite{spike}.

\section{The Sensitivity of GLAST}

\begin{figure}

\resizebox{8cm}{!}{\includegraphics{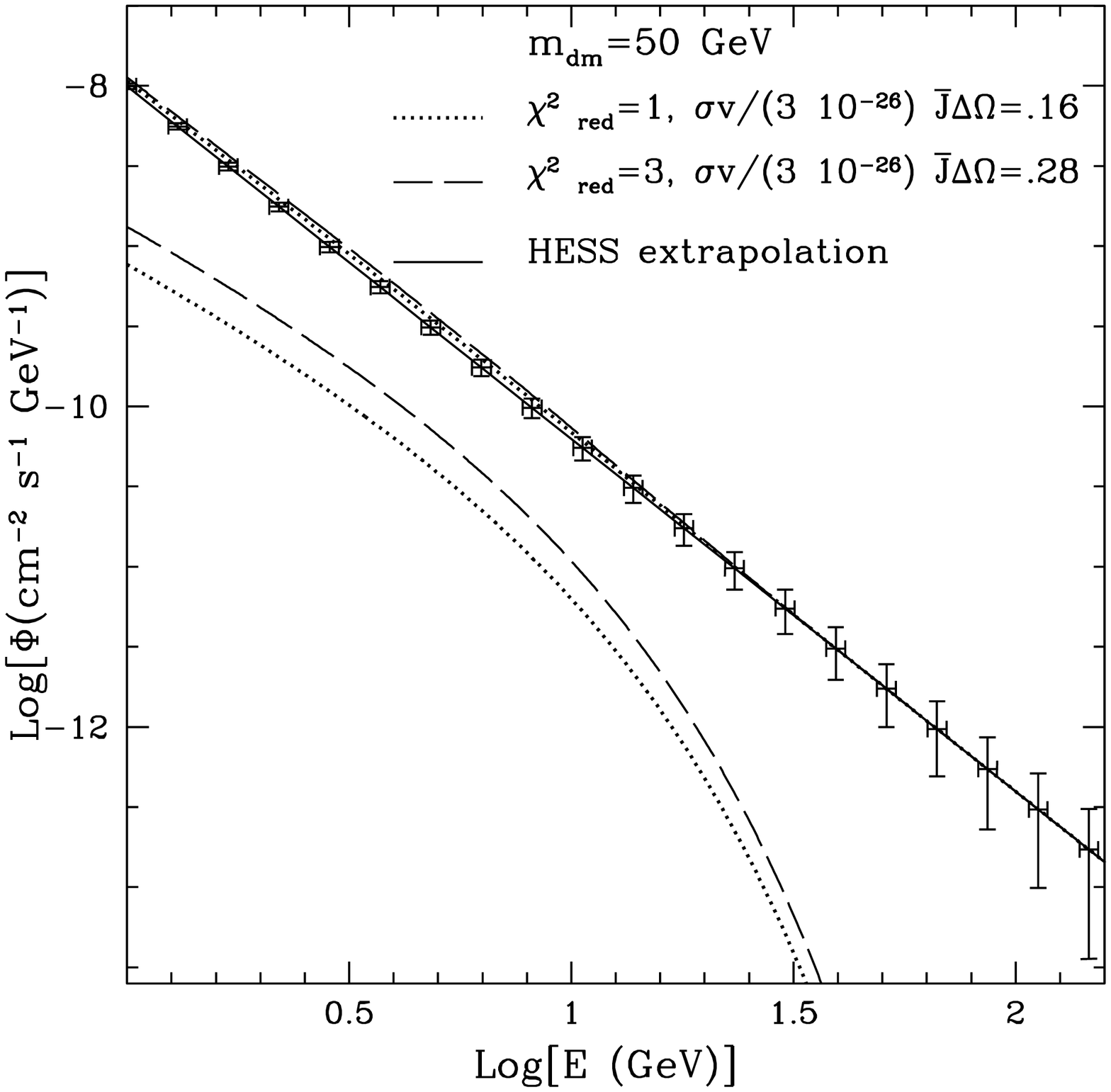}}
\resizebox{8cm}{!}{\includegraphics{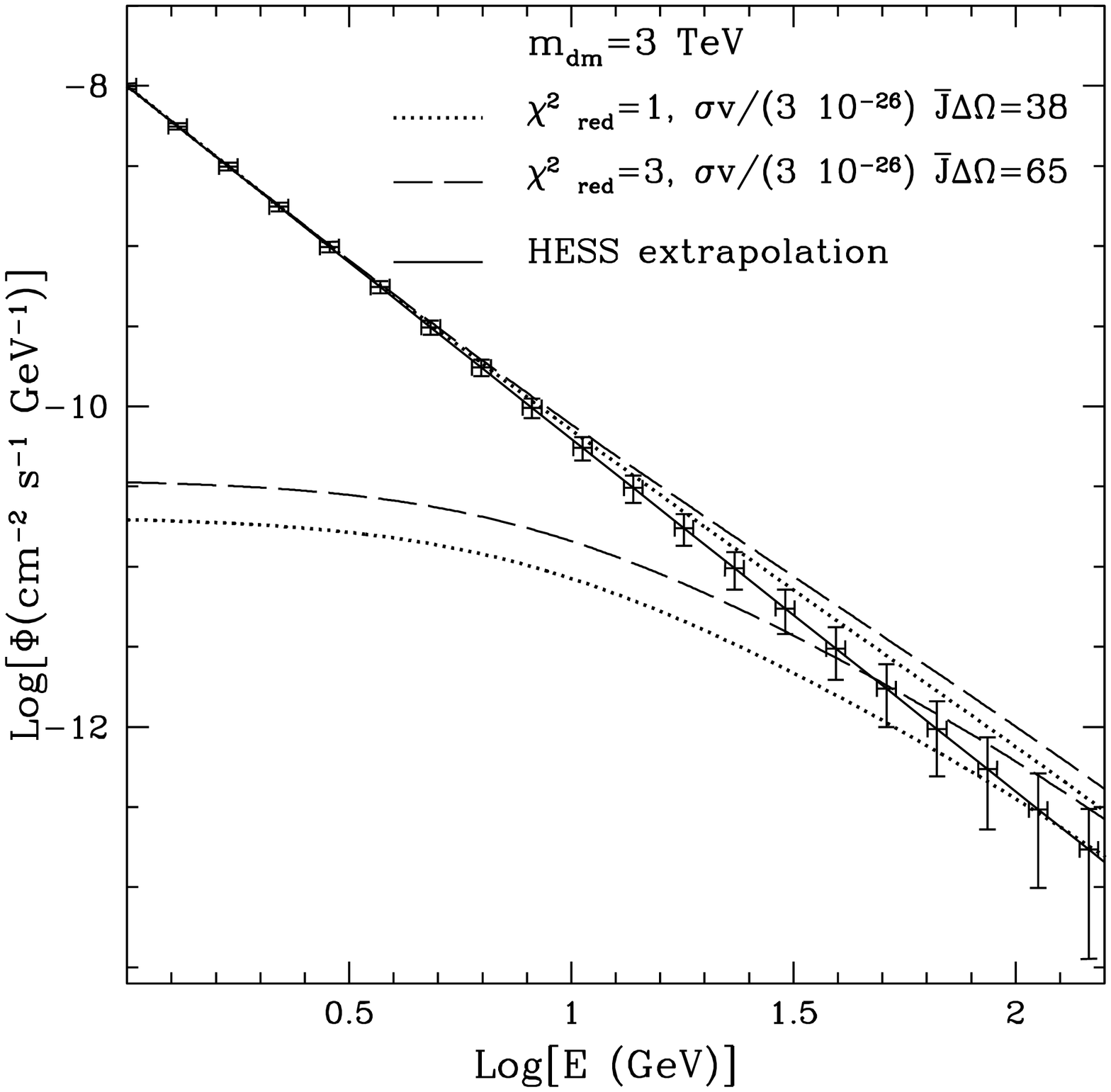}}
\caption{The minimum flux of gamma-rays from dark matter in the galactic center needed to be detected by GLAST, assuming a power-law background extrapolated from the HESS data (see Fig.~\ref{astro}). The two frames correspond to two dark matter masses: 50 GeV (left) and 3 TeV (right). For a 50 GeV WIMP, the quantity $(<\sigma v>/(3 \times 10^{-26}$cm$^3$/s))$ \times J(\Delta \Omega) \Delta \Omega$ needs to be larger than $\sim$0.2 to be identified by GLAST. For a 3 TeV WIMP, this quantity must be larger $\sim$50 to be detected. For GLAST, we have used an effective area $\times$ exposure time of 1 m$^2$ yr.}
\label{glast}


\resizebox{10cm}{!}{\includegraphics{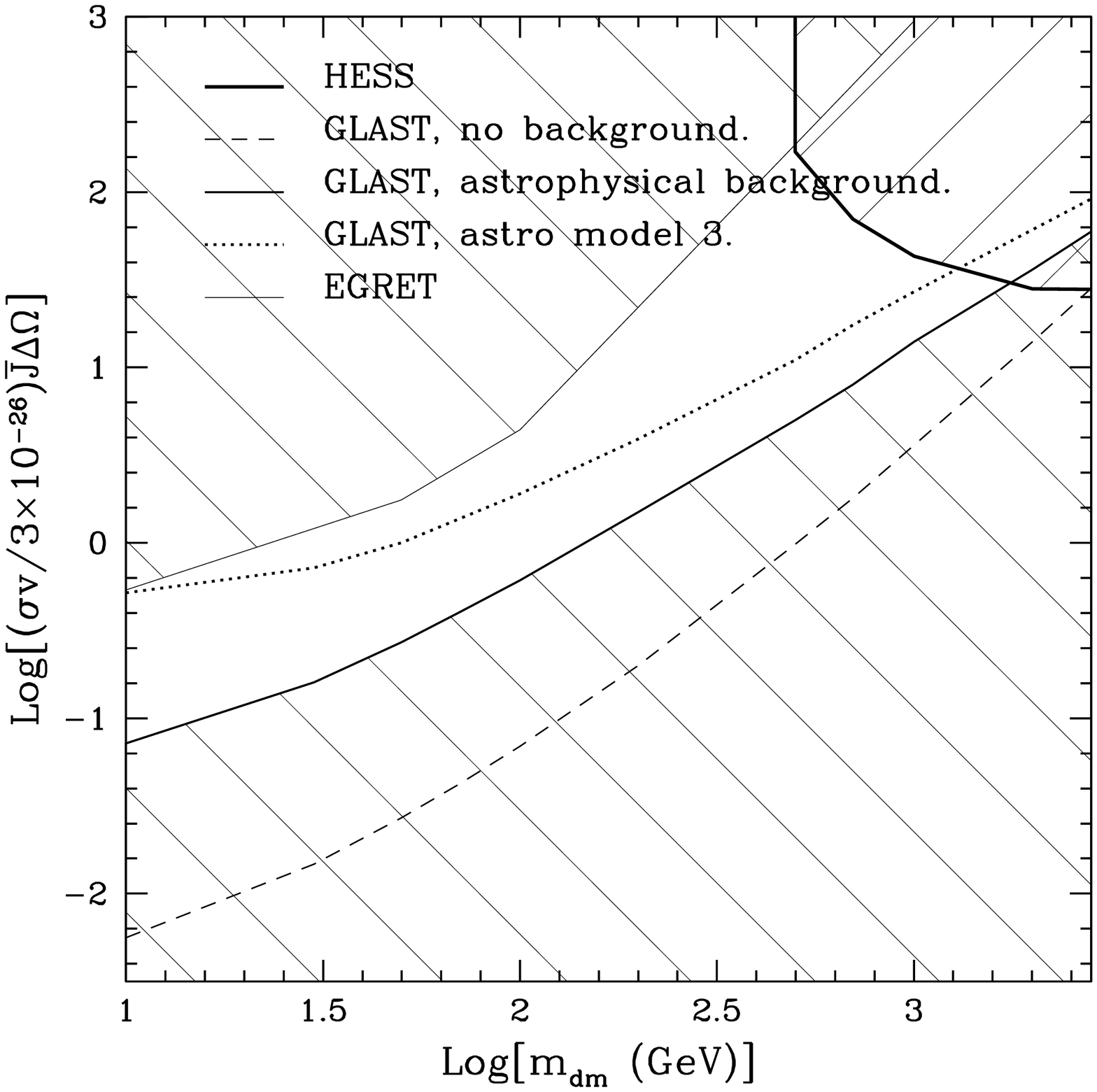}}
\caption{The range of annihilation rates and WIMP masses excluded by HESS and/or EGRET and within the reach of GLAST. The models which remain potentially observable by GLAST are those in the non-hatched region in the middle of the figure. The upper-right and upper-left hatched regions are currently excluded by HESS or EGRET \cite{dingus}, respectively.  The lower hatched region represents the models which will be undetectable with GLAST, considering the astrophysical background extrapolated from HESS. If instead we considered the astrophysical background model 3 of Fig.~\ref{astro}, however, this region extends up to the dotted line. Also shown, for comparison, is the reach of GLAST if no astrophysical source were present (dashed line).}
\label{glasthess}
\end{figure}

The satellite-based gamma-ray telescope GLAST, scheduled to begin its five year mission in 2007, will be sensitive to gamma-rays of energy between approximately 100 MeV and 300 GeV. With an effective area and angular resolution on the order of $10^4$ cm$^2$ and 0.1$^{\circ}$, respectively, GLAST will be substantially more sensitive to dark matter annihilation than its predecessor, EGRET \cite{glast,dingus,buckley}.

In Fig.~\ref{glast}, we show the ability of GLAST to measure the gamma-ray spectrum from the galactic center, and potentially identify a component from dark matter. The error bars shown are projected for an astrophysical background only (assuming a power-law background extrapolated from the HESS data). We find that for a 50 GeV WIMP, the $\chi^2$ of the fit to the extrapolated power law becomes poor (possibly enabling for the identification of the contribution from dark matter) for $(<\sigma v>/(3\times 10^{-26} \,\rm{cm}^3/\rm{s})) \times \bar{J}(\Delta \Omega)\Delta \Omega$ larger than approximately 0.2. For a 3 TeV WIMP, the fit becomes poor only for $(<\sigma v>/(3\times 10^{-26} \,\rm{cm}^3/\rm{s}))\times \bar{J}(\Delta \Omega)\Delta \Omega$ larger than approximately 50.

In Fig.~\ref{glasthess}, we compare the sensitivity of GLAST to the constraints of HESS. The upper-right hatched region is currently excluded by HESS. The upper-left hatched region has been excluded by EGRET observations of the galactic center \cite{dingus}.  The lower hatched region represents the models which will be undetectable with GLAST, considering the astrophysical background extrapolated from HESS (see Fig.~\ref{glast}). If instead we considered the astrophysical background model 3 of Fig.~\ref{astro}, however, such a detection becomes more difficult and the undetectable region extends up to the dotted line. 

We can see that for a 50 GeV WIMP, for example, there is only a range of a factor of about 3 in the annihilation rate ($<\sigma v> \bar{J}(\Delta \Omega)\Delta \Omega$) over which GLAST would potentially be able to detect dark matter annihilation radiation. For a heavier WIMP, this range becomes larger, spanning about a factor of 20 for a 300 GeV WIMP, for example. For a WIMP heavier than about 1.7 TeV, however, the annihilation rate is always either excluded by HESS, or is below the sensitivity of GLAST. 

Also shown in Fig.~\ref{glasthess} is the reach of GLAST if no astrophysical source were present in the galactic center region. It is clear that a substantial fraction of the range of models which were previously thought to be within the reach of GLAST are no longer accessible given the presence of the astrophysical source.

The approach we are taking here should be taken as a idealized treatment for GLAST, {\it ie.} a best-case-scenario. Because we do not know in advance the shape of the astrophysical background in the 1--300 GeV range, it may be difficult to determine with confidence whether dark matter annihilation is present in the spectrum of GLAST. For this reason, the range of annihilation rates which could be identified by GLAST may be even more narrow than is shown in Fig.~\ref{glasthess}.

\section{Gamma-Ray Lines}

In addition to generating continuum gamma-rays through the decays of quarks, leptons, Higgs bosons or gauge bosons, dark matter particles can produce gamma-rays directly, leading to monoenergetic spectral signatures. If a gamma-ray line could be identified, it would constitute a ``smoking gun'' for dark matter.

Neutralinos, for example, can annihilate directly to $\gamma \gamma$ \cite{gg} or $\gamma Z$ \cite{gz} through a variety of loop diagrams. These final states lead to gamma-rays lines with energies of $E_{\gamma}=m_{\rm{dm}}$ and $E_{\gamma}=m_{\rm{dm}}(1-m^2_Z/4m^2_{\rm{dm}})$, respectively. Such photons are produced in only a very small fraction of neutralino annihilations, however. The largest neutralino annihilation cross sections to $\gamma \gamma$ and $\gamma Z$ are about $10^{-28}$ cm$^3$/s, although much smaller values are more typical~\cite{buckley}.

The number of line events detected by GLAST or an ACT from dark matter annihilations in the galactic center is given by:

\begin{equation}
N_{\rm{line}} \approx 1.9 \times 10^{-15} \,C \bigg(\frac{<\sigma v>_{\rm{line}}}{10^{-29} \,\rm{cm}^3/\rm{s}}\bigg)  \bigg(\frac{1 \, \rm{TeV}}{m_{\rm{dm}}}\bigg)^2 J(\Delta \Omega) \Delta \Omega \, A_{\rm{eff}}\, T,
\label{lineflux}
\end{equation}
where $C=1,2$ for annihilations to $\gamma Z$ and $\gamma \gamma$, respectively.  $A_{\rm{eff}}$ and $T$ are the effective area (in cm$^{2}$) and exposure time (in seconds) of the experiment.

For GLAST to detect the presence of a gamma-ray line, the signal must overcome the astrophysical background from the source in the galactic center region, integrated over the energy resolution of the experiment. For a positive detection, we adopt the standard discovery criterion: $N_{\rm{line}}/\sqrt{N_{\rm{bg}}} > 5$.

For ACTs such as HESS or MAGIC, an additional background must be considered. Hadronic cosmic rays whose showers are misidentified as electromagnetic generate a background spectrum in ACTs of \cite{buckley}
\begin{equation}
\frac{dN_{\rm{had}}}{dE_{\rm{had}}} \approx 3 \epsilon \times E(\rm{GeV})^{-2.7} \, \rm{GeV}^{-1}\, \rm{cm}^{-2} \, \rm{s}^{-1} \, \rm{sr}^{-1},
\end{equation}
where $\epsilon$ is the fraction of hadronic showers which are misidentified and thus not rejected. For modern ACTs, $\epsilon \sim $1\%.

\begin{figure}

\resizebox{8cm}{!}{\includegraphics{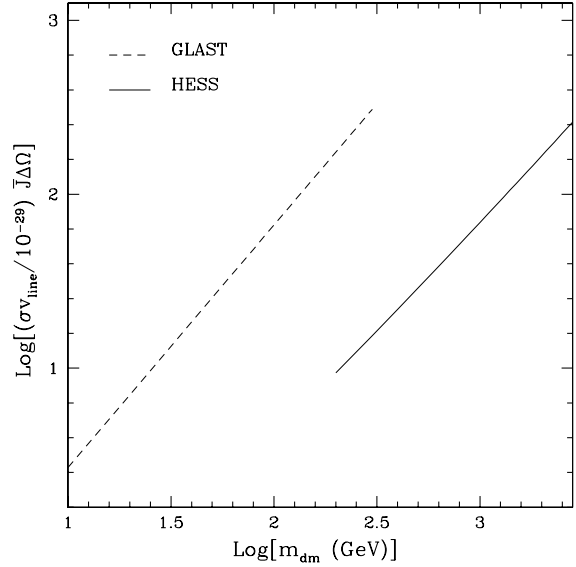}}
\caption{The range of annihilation rates to gamma-ray lines and WIMP masses within the (5$\sigma$) reach of GLAST or HESS. The cross section shown in the y-axis is that to $\gamma \gamma$. If a $\gamma Z$ line is being considered, the cross section must be larger by a factor of 2 to be detected.}
\label{glasthesslines}
\end{figure}

In Fig.~\ref{glasthesslines}, we plot the reach of GLAST and HESS in detecting gamma-ray lines from the galactic center. Roughly speaking, we find that these experiment will be capable of identifying a line feature if the quantity $(\sigma v/10^{-29}) \bar{J}(\Delta \Omega)\Delta \Omega$ is larger than $\sim$10--100. We have estimated energy resolutions of 10\% and 15\% for GLAST and HESS, respectively, and considered 80 hours of observation with HESS. Again, we have considered 1 m$^2$ yr exposure for GLAST.

\begin{figure}

\resizebox{8cm}{!}{\includegraphics{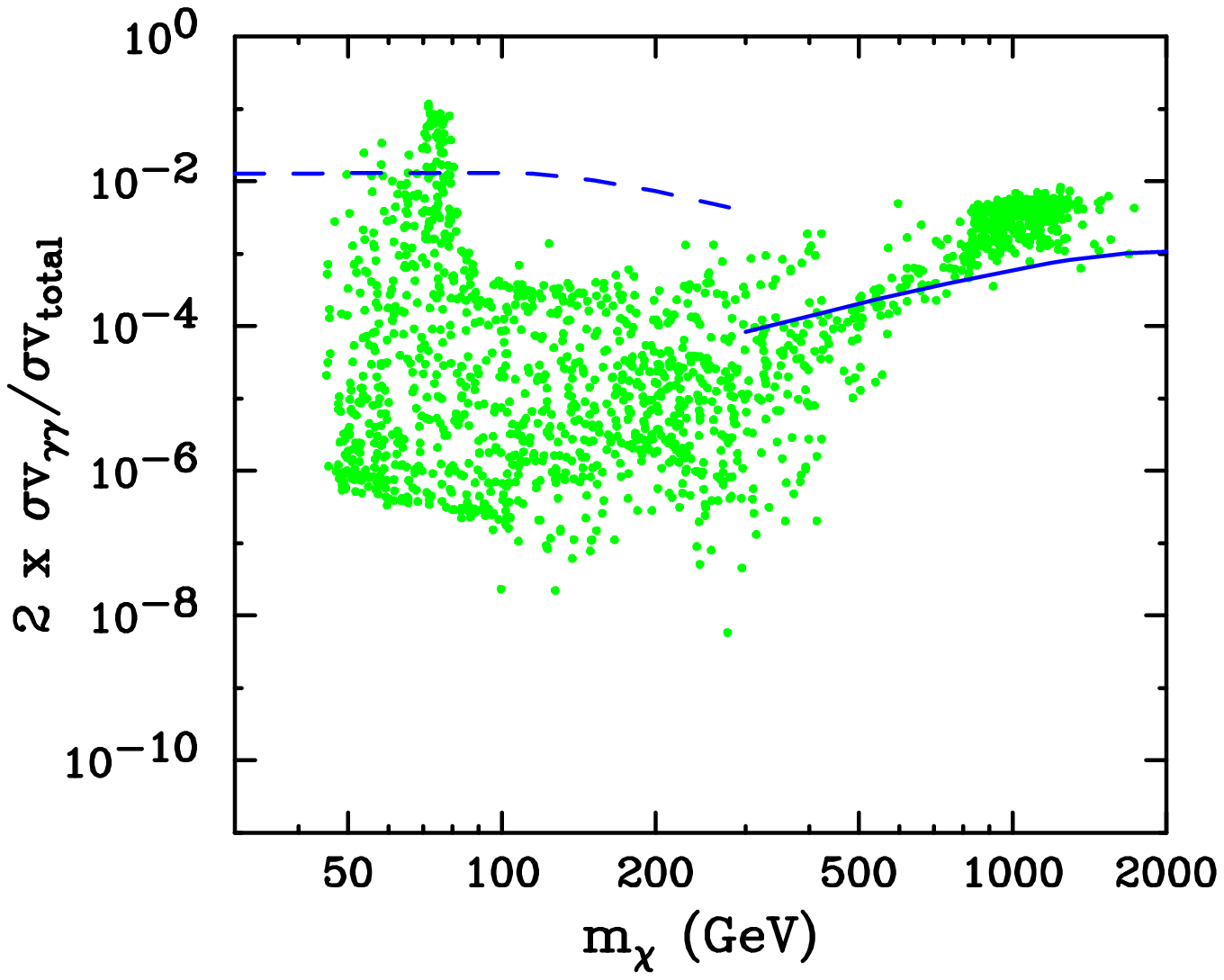}}
\resizebox{8cm}{!}{\includegraphics{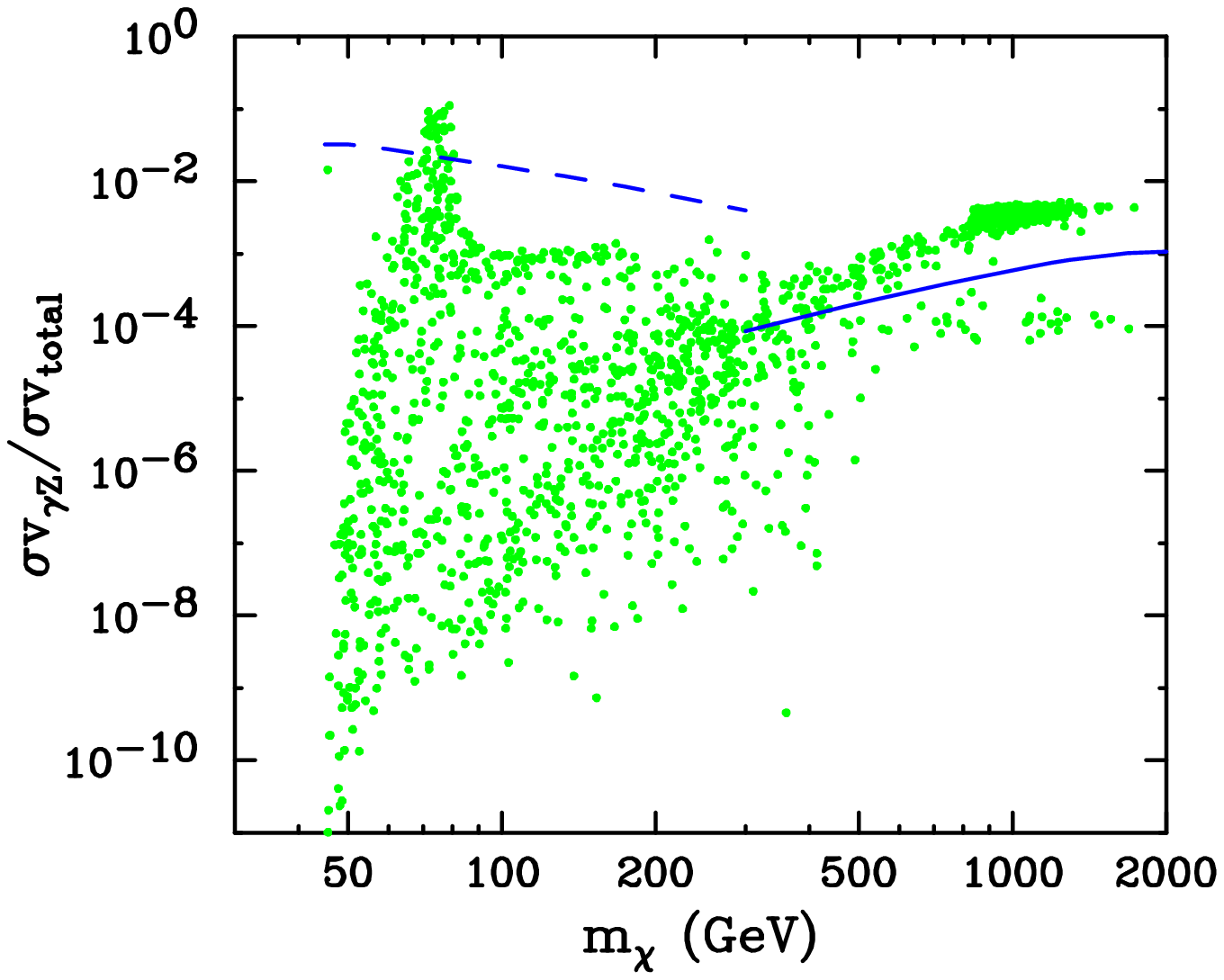}}
\caption{The neutralino annihilation cross section to $\gamma \gamma$ (left) and $\gamma Z$ (right) over the total annihilation cross section (at low velocity) for a range of randomly selected supersymmetric parameters. In order for a gamma-ray line to be detectable by GLAST or HESS, the annihilation rate must be small enough to not exceed the HESS and EGRET constraints on continuum emission and large enough to generate an observable line signal. For the points shown above the lines, there exists a range of $\bar{J}(\Delta \Omega)\Delta \Omega$ which satisfies this criterion. Therefore, only those points above the dashed and solid lines are potentially capable of producing observable line signatures in GLAST and HESS, respectively.}
\label{sigmalines}
\end{figure}

Comparing the results of Figs.~\ref{glasthess} and~\ref{glasthesslines}, we see that the ability of GLAST or HESS to detect line emission is constrained by the existing observations of HESS and EGRET. Note, however, that the HESS and EGRET constraints shown in Fig.~\ref{glasthess} constrain the total annihilation cross section rather than the annihilation to $\gamma \gamma$ or $\gamma Z$ lines. To make a comparison of these two quantities, we must consider a specific dark matter model, such as supersymmetric neutralinos. In Fig.~\ref{sigmalines}, we show the ratio of the neutralino's annihilation cross section to lines to the total annihilation cross section (at low velocity). The points shown represent randomly selected parameter values within the Minimal Supersymmetric Standard Model (MSSM), calculated with the DarkSUSY program~\cite{darksusy}. Each point shown generates the observed cold dark matter density and is consistent with all collider constraints.

It is clear that if the annihilation cross section to lines is very small compared to the total annihilation cross section, then limits from EGRET and HESS on the continuum emission from the galactic center will make observing lines from the galactic center difficult or impossible. In Fig.~\ref{sigmalines}, all points below the dashed and solid lines will not be observed as lines by GLAST and HESS, respectively, {\it for any halo profile} ({\it ie.} any value of $\bar{J}\Delta \Omega$). Those points above the lines could potentially produce observable lines in these experiments. Notice, however, that these points are not far above the lines and, therefore, for such a detection to be made, the value of $\bar{J}(\Delta \Omega) \Delta \Omega$ must be within the narrow range needed to both evade the continuum constraints of EGRET and HESS and generate a bright enough line signal. This leads us to the conclusion that, although not impossible, a line detection from the galactic center region by GLAST or HESS is very unlikely, at least for the case of neutralino dark matter.

\section{Summary and Conclusions}

A bright source of gamma-rays has recently discovered in the region of the galactic center. In this article, we have explored the impact of this source on the prospects for detecting gamma-rays generated in dark matter annihilations with experiments such as GLAST. We find that this source constitutes a significant background that will make efforts to detect dark matter in the galactic center very difficult.

Once this background is taken into account, and constraints from  EGRET and HESS are considered, only a narrow range of annihilation rates is potentially within the reach of GLAST. Given these considerations, very heavy WIMPs ($m_{\rm{dm}}>$1.7 TeV) in the galactic center will not be observable by GLAST, for any annihilation rate consistent with current observations.

All hope should not be lost, however. Quite interestingly, the range which is still observable by GLAST includes the expected spectrum for a thermally generated ($\sigma v \approx 3 \times 10^{-26}$ cm$^3$/s), 20-200 GeV WIMP with an NFW-like halo profile. Heavier WIMPs, up to 1--1.7 TeV, would be within the accessible range if a Moore {\it et al.} profile were present. So, despite the challenges faced by GLAST, very attractive dark matter scenarios remain within its reach.

We have also studied the prospects for detecting monoenergetic gamma-ray lines from dark matter annihilations with GLAST or HESS. Here we find the prospects to be very poor. For most neutralino models, for example, no range of halo models can produce an observable line in GLAST or HESS without exceeding the continuum spectrum observed by EGRET and HESS. In those few models which produce comparatively bright lines, only very narrow ranges of halo models would produce an observable line given current constraints and backgrounds.

In light of these considerable challenges, dark matter searches using gamma-rays may be more fruitful in regions other than the galactic center. Dwarf spheroidal companion galaxies to the Milky Way are among the most promising~\cite{dwarf}. Also potentially interesting are galaxies external to the Milky Way \cite{external}, or dark matter density spikes around intermediate mass black holes \cite{imbh}.

\bigskip

We would like to thank Glennys Farrar for providing many helpful comments. This work has been supported by the US Department of Energy and by NASA grant NAG5-10842.

\end{document}